# Effects of CdCl$_2$ treatment on the local electronic properties of polycrystalline CdTe measured with photoemission electron microscopy

Morgann Berg[1,2], Jason M. Kephart[3], Walajabad S. Sampath[3], Taisuke Ohta[1,2], Calvin Chan[1]

[1]Sandia National Laboratories, Albuquerque, New Mexico, 87185, USA
[2]Center for Integrated Nanotechnologies, Albuquerque, New Mexico, 87185, USA
[3]Department of Mechanical Engineering, Colorado State University, Fort Collins, Colorado, 80523, USA

*Abstract —* To investigate the effects of CdCl$_2$ treatment on the local electronic properties of polycrystalline CdTe films, we conducted a photoemission electron microscopy (PEEM) study of polished surfaces of CdTe films in superstrate configuration, with and without CdCl$_2$ treatment. From photoemission intensity images, we observed the tendency for individual exposed grain interiors to vary in photoemission intensity, regardless of whether or not films received CdCl$_2$ treatment. Additionally, grain boundaries develop contrast in photoemission intensity images different from grain interiors after an air exposure step, similar to observations of activated grain boundaries using scanning Kelvin probe force microscopy studies. These results suggest that work function varies locally, from one grain interior to another, as well as between grain boundaries and grain interiors.

## I. Introduction

CdTe thin-film photovoltaics generate more than 2 GW annually, with research cell efficiencies (21.5%) rivaling those of multicrystalline Si. As such, CdTe-based solar cells are a serious alternative to silicon-based photovoltaics for commercial production and deployment. A key component of manufacturing efficient CdTe photovoltaic devices is an activation process where the polycrystalline CdTe absorber layer is treated with CdCl$_2$ [1]. The effect of CdCl$_2$ treatment on the electronic properties of CdTe-based devices has been extensively studied but is not completely understood.

One consequence of CdCl$_2$ treatment is a change in the microstructure of the CdTe layer, resulting in larger and more uniform grains, and increased minority carrier lifetimes [1]. While fewer grain boundaries are thought to reduce scattering and recombination of photogenerated carriers, the performance of single crystal CdTe photovoltaics is known to be inferior to that of polycrystalline films.

This contradiction is often explained by the modification of the electronic properties at grain boundaries during CdCl$_2$ treatment. The prevailing description is that Cl from the CdCl$_2$ treatment segregates to grain boundaries in p-type CdTe, depleting or inverting them with respect to grain interiors. Since the formation of junctions between grain interiors and grain boundaries are expected to conduct charges to the electrodes more efficiently, individual grain boundary properties have been a subject of intense study, which have been measured using an array of techniques including scanning probe microscopy [2]–[3], electron beam-induced current [4], time-resolved photoluminescence [5], and time-of-flight secondary ion mass spectroscopy (ToF-SIMS) [6].

For thin-film junctions, photoemission spectroscopy is invaluable for directly probing variation in electronic properties. For example, ultraviolet photoelectron spectroscopy (UPS) is routinely used to measure Fermi level positions, estimate doping levels, and determine relative band alignment among thin-film components in semiconductor devices. However, because UPS is an area-averaged technique, lateral variation of electronic properties smaller than the beam spot size (typically 1 mm to tens of microns) cannot be resolved, rendering measurement of local junctions (e.g. between grain interiors and grain boundaries in CdTe thin films) inaccessible.

In the work reported here, we used photoemission electron microscopy (PEEM) to image the effect of CdCl$_2$ treatment on the local properties of superstrate CdTe films, exposed at the CdTe/rear contact interface. To expose the grain interiors for PEEM measurements, the CdTe films were mechanically polished. To isolate the influence of CdCl$_2$ vapor treatment on CdTe surfaces, we used low-energy ion desorption to remove surface oxides and contaminants from polished films [7]. We observed grain domains and saw little change in microstructure between CdCl$_2$-treated and untreated films. From changes in photoemission intensity we deduced that individual grains vary in work function for both CdCl$_2$-treated and untreated films. Surprisingly, our results indicate that CdCl$_2$-treatment alone does not modify grain boundary properties in a manner consistent with previous reports [2]–[3]. After purposefully exposing the CdCl$_2$-treated CdTe film to air, the sample surface develops image contrast at grain boundaries in photoemission intensity images. These findings suggest a synergistic relationship between CdCl$_2$ treatment and oxygen incorporation changes grain boundary properties.

## II. Experimental Details

CdTe samples were fabricated using closed-space sublimation (CSS) at Colorado State University's Advanced Research Deposition System [8]. Films were deposited on commercial NSG TEC 10 superstrates, which feature standard soda lime glass coated with a transparent conducting oxide (TCO). The TCO, consisting of a SnO$_2$/SiO$_2$/SnO$_2$:F tri-layer, was cleaned using standard rinses and an isopropyl alcohol wash, followed by a plasma cleaning process [9]. A 100-nm-thick, $Mg_{0.23}Zn_{0.77}O$ (MZO) window layer was sputtered onto the TCO. 5 μm CdTe was then deposited directly onto the MZO. During CSS growth and $CdCl_2$ treatment, substrate temperatures were in the range of 425–500°C. Source temperatures were 435–610°C. To remove topographical artifacts in PEEM measurement and to expose the relevant CdTe grain interiors, samples were mechanically polished to ~3 μm thickness using a 1 μm diamond polish. Samples were subsequently sputter-cleaned with 50 eV $Ar^+$ ions for 10–20 min with a fluence of ~0.1-0.15 μA•$cm^{-2}$ to remove surface contaminants [7]. This low energy sputtering process is conventionally referred to as low-energy ion desorption within the community. We verified the composition of the CdTe surface before and after surface modification using x-ray photoelectron spectroscopy (XPS). XPS survey scans were performed using a Mg $K_\alpha$ x-ray source and an Omicron EAC 2000 electron analyzer, operating at a pass energy of $E_p$ = 50 eV. XPS core-level scans were also obtained by operating the electron analyzer with a pass energy of $E_p$ = 20 eV. After sputter-cleaning and XPS measurement, samples were transferred to PEEM without additional exposure to air by the use of an inert environment sample transfer system and an inert gas ($N_2$) glove box. To examine the impact of oxygen on CdTe processing, the $CdCl_2$-treated film was exposed to air for 30 minutes, reintroduced into ultrahigh vacuum (UHV), and measured again.

Figure 1 provides visual representations of the superstrate CdTe films (Fig. 1a) and PEEM electronic property measurement. Fig. 1b identifies physical grains and grain boundaries in the polished surface. Fig. 1c illustrates our designations for electronic property variation that occurs between one grain surface to another (grain-to-grain), and between a grain surface and grain boundary (grain-to-boundary).

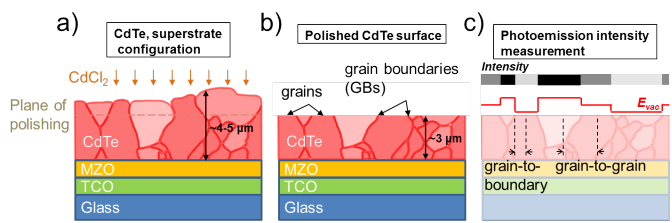

Fig. 1. (a) Schematic of a CdTe sample in superstrate configuration, showing the plane of polishing to expose the grain interiors. (b) Grain interiors and grain boundaries (GB) are illustrated to define grain-to-grain and grain-to-boundary designations (c) for electronic structure variation measured with PEEM. Fig. 1c also shows hypothetical variation of the local photoemission intensity due to variation of the vacuum level ($E_{VAC}$) plotted in red. Dark regions in the photoemission intensity profile of Fig. 1c denote regions with relatively low photoemission intensity and light regions correspond to regions with relatively high photoemission intensity.

Regions with relatively higher vacuum level in Fig. 1c correspond to regions of lower photoemission intensity, represented by dark regions. Similarly, light regions in the photoemission intensity profile in Fig. 1c, corresponding to higher photoemission intensity, are correlated to regions with lower vacuum level. We further discuss the relationship between photoemission intensity variation and variation in electronic properties later in this section, following descriptions of PEEM measurement.

PEEM measurements were conducted in a LEEM-III system (Elmitec Elektronenmikroskopie GmbH) equipped with a hemispherical electron energy analyzer and coupled to a tunable DUV light source composed of a pressurized Xe lamp (Energetiq, EQ-1500 LDLS), a Czerny–Turner monochromator (Acton research, SP2150), and refocusing optics. The spectral width of the DUV light was set to 50-100 meV throughout the wavelength range used for the measurement ($\lambda$ = 175–350 nm, $h\nu \cong$ 3.6–7 eV). The field of view (FOV) for all photoemission images was 48 μm at 600 pixels, corresponding to a pixel resolution of ~80 nm/pixel.

Using a fixed photon wavelength, $\lambda$ = 190 nm ($h\nu \cong$ 6.5 eV) as was the case for this study, we acquire photoemission spectra (PES) at each pixel in a PEEM image by sweeping the voltage offset (start voltage, $V_s$) of the electron kinetic energy spectrum with respect to the energy window of the analyzer. For this study, photoemission intensity images were collected at each $V_s$ value, with a step size of $\Delta V_s$=10 meV. This yields a photoemission intensity versus electron kinetic energy profile at each pixel. Similar to conventional analysis of photoelectron spectra, onsets and cutoffs of the spectra in energy correspond to locations in energy of the vacuum level ($E_{VAC}$) and the valence band edge ($E_{VBE}$).

Changes in photoemission intensity are not necessarily an indicator of local electronic property variation. Both local variation in the photoemission cross section and local variation in the work function can result in variation of the photoemission intensity. Ruling out the contribution of photoemission cross section requires extracting local vacuum level data from PES. However, in the absence of significant variation in the photoemission cross section for a given photoexcitation wavelength, regions of lower relative work function typically correspond to higher relative photoemission intensity. This relationship is reflected in the hypothetical variation of the vacuum level and photoemission intensity presented in Fig. 1c, where the areal variation in the $E_{VAC}$ and intensity appear to be inversely correlated.

Assuming the Fermi level is aligned throughout the sample, differences in the local vacuum level may be attributed to local variation in the relative work function. Thus, local variation of photoemission intensity observed in PEEM images obtained near the vacuum level of PES can provide a qualitative indicator of local work function variation in the sample surface. We proceed to discuss results assuming that the local photoemission cross section does not vary significantly, and that photoemission intensity variation is an indicator of work function variation.

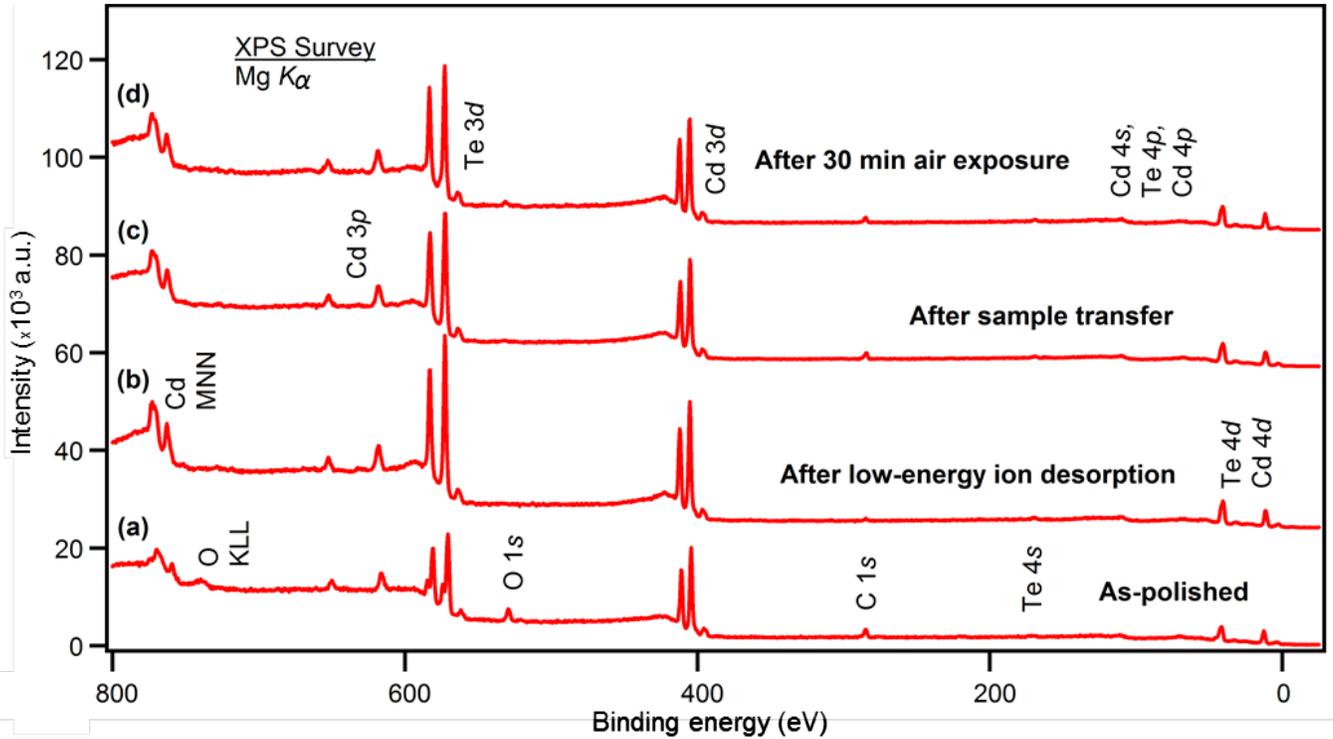

Fig. 2 XPS survey spectrum of the CdCl2-treated CdTe surface after polishing (a), showing O and C in addition to Cd and Te. After a low-energy ion desorption step (b) the O signal is undetectable within the noise level, the C signal is dramatically reduced, and intensity from Cd and Te core peaks are increased. Sample transfer using the inert gas glove box (c) introduces carbon, while the O signal remains below the noise level. 30 min of air exposure (d) results in a small increase in the C and O signals. Spectra (b)-(d) are offset for clarity.

## III. RESULTS & DISCUSSION

Figure 2 shows XPS spectra of the CdCl$_2$-treated CdTe surface after various sample preparation steps. For the as-polished sample (Fig. 2a), in addition to a pronounced O1$s$ and O KLL signal, the Te3$d$ profile shows additional peaks shifted to higher binding energy. These peaks are associated with TeO$_x$ formation. These signatures, associated with the presence of surface oxygen, are undetectable within the noise level after brief sputtering of the surface with 50 eV Ar$^+$ ions (Fig. 2b). The carbon (C1$s$) signal is almost completely removed after sample cleaning. Sample transfer using an inert gas glove box (Fig. 2c) introduces surface carbon prior to PEEM measurement, and air exposure (Fig. 2d) results in increase of the O1$s$ and C1$s$ signals. Air exposure also results in a small TeO$_x$ component that is apparent in fits of the Te3$d$ profile and in core-level scans.

To quantify the amount of carbon introduced during sample transfer through the inert gas glove box, 30 min air exposure, and PEEM measurement, we used a graphene monolayer (1ML) on SiO$_2$ as a reference to quantify the carbon coverage of films. After polishing, the C1$s$ signal is comparable to one-third of a monolayer (1/3 ML) of carbon. 30 minutes of air exposure produces a C1$s$ signal comparable to 1/5 ML of carbon. The highest C1$s$ signal comparable to 2/5 ML of carbon is observed for the CdTe surface after PEEM measurement (not shown). The magnitude of the C1$s$ signal after PEEM measurement is likely due to transferring the sample twice through the inert gas glove box, from XPS to PEEM and then from PEEM back to XPS.

Cd:Te compositional ratios were not greatly impacted by mechanical polishing or ion-desorption steps. Cd:Te compositional ratios were ~5:6 before and after ion desorption of the polished, untreated CdTe surface, and ~11:10 for the polished, CdCl$_2$-treated sample. The Cd:Te ratio of the CdCl$_2$-treated sample before polishing was slightly higher, ~12:10. These compositional ratios are close to 1:1 as expected from the 1:1 stoichiometry of CdTe. Low-energy ion desorption broadened Cd3$d_{5/2}$ and Te3$d_{5/2}$ peaks but introduced no additional asymmetry, and no Cl2$p$ signals were detected from the surface. These results are consistent with Ref. [7].

Figure 3 displays PEEM photoemission intensity images (FOV = 48 μm) of CdCl$_2$-treated and untreated samples. To highlight local variations in the vacuum level within individual images, the potential offset (start voltage) of the electron kinetic energy at which each image was acquired was close to the vacuum level edge of PES, $V_s \sim E_{VAC}$. Specifically, the photoemission intensity image in Fig. 3a was obtained at a start voltage, $V_s = -0.21$ V. For Fig. 3b a start voltage of 0.09 V was used, and for Fig. 3c $V_s = 0.02$ V. The color scales were adjusted to enhance local contrast for each image, individually. Although the photoemission intensity scales in Fig. 3 are not normalized, the scale bar on the left of Fig. 3 shows that dark and light regions correspond to regions of low and high photoemission intensity, respectively. We note that Fig. 3a displays field aberrations in the lower left quadrant that resulted from a local sample/holder contact problem.

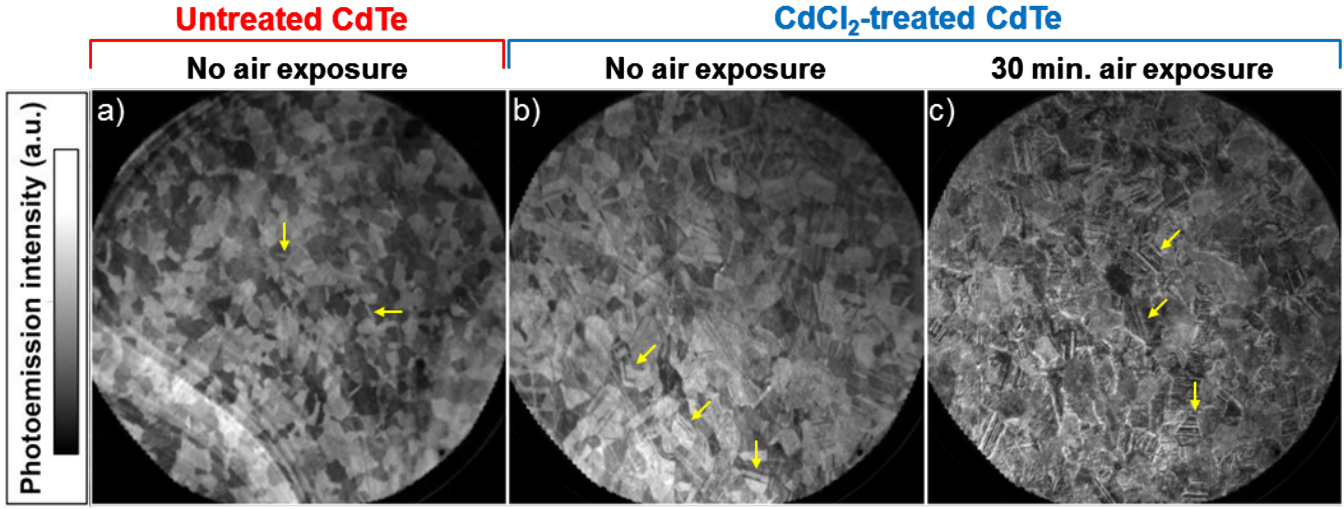

Fig. 3. Photoemission intensity images of (a) untreated CdTe, and CdCl$_2$-treated CdTe, before (b) and after (c) air exposure. Images were obtained at a start voltage value near the vacuum level edge of PES, $V_s \sim E_{VAC}$. In particular, for Fig. 3a $V_s = -0.21$ V, for Fig. 3b $V_s = 0.09$ V, and for Fig. 3c $V_s = 0.02$ V. PES were acquired using $\lambda = 190$ nm ($h\nu \sim 6.5$ eV) light. Yellow arrows identify grain domains that contain what appears to be planar defects or twin boundaries, consistent with those observed in Ref. [12]. The field of view for all images is 48 μm.

Contrast in all photoemission intensity images in Fig. 3 varies amongst domains on the scale of microns. These micron-scale domains correspond to the microcrystalline CdTe grains as identified in other microscopy studies of polycrystalline CdTe films [10]. The domain sizes are consistent with grain sizes of polycrystalline CdTe films grown at higher temperatures [11]. There are also narrow domains in Fig. 3 running parallel to each other with abrupt interfaces between them (yellow arrows). These abrupt interfaces bear a resemblance to microscopic observations of planar defects and twin boundaries [12]. Polish marks are also seen in Figs. 3b,c as narrow lines with relatively low photoemission intensity that are continuous as they cross multiple grain domains. Grains in CdCl$_2$-treated CdTe (Fig. 3b) seemed to be 1-2 times larger than grains in untreated CdTe (3a). The change in grain size we observed with PEEM is consistent with observations of grain recrystallization after CdCl$_2$ treatment for CdTe films grown at higher temperature [13]. Though grain size is recognized as an important parameter impacting the performance of CdTe as an absorber layer [14]–[15], Ref. [13] demonstrated that CdCl$_2$ treatment does not greatly impact the grain size, grain orientation, nor the distribution and structure of grain boundaries in CSS CdTe films grown at high temperature. Thus, the authors of Ref. [13] concluded that improvements in efficiencies observed after CdCl$_2$ treatment of films grown at high temperature were not driven by changes in microstructure.

All films display contrast in the photoemission intensity that varies on the scale of grains (grain-to-grain variation). Photoemission intensity images of the untreated (Fig. 3a) and CdCl$_2$-treated (Fig. 3b) CdTe show similar grain-to-grain variation. Thus, the impact of CdCl$_2$ vapor treatment is not readily apparent from photoemission intensity images. The presence of grain-to-grain contrast in all films we studied suggest that there may be an inherent grain-to-grain work function variation. Though grain-to-grain non-uniformity is not an entirely new finding [3], [6], it has been less explored as a factor impacting local junction formation between CdTe and neighboring buffer layers and electrodes. It should be expected that carrier drift and diffusion could vary significantly depending on the orientation of CdTe crystallites, as suggested by our spatially varying work function measurements.

Surprisingly, after air exposure of the CdCl$_2$-treated film, contrast at the interfaces between grain domains (grain boundaries) appears in the photoemission intensity image. This contrast is visible as narrow, bright lines decorating the edges of grain domains in Fig. 3c. The lack of grain-to-boundary variation in the CdCl$_2$-treated film (Fig. 3b) and its appearance after exposure to air (Fig. 3c) suggests that, on its own, CdCl$_2$ vapor treatment does not individuate grain boundary properties from those of grain interiors. Observations of increased surface potential at grain boundaries [3] and increased current collection [4] have been previously attributed to segregation of Cl to grain boundaries following CdCl$_2$ treatment. Our findings instead suggest that incorporation of oxygen into a CdCl$_2$-treated film induces grain-to-boundary property variation.

## VI. Conclusion

In conclusion, initial photoemission electron microscopy measurements of the CdTe surface suggest that electronic properties vary from one grain interior to another, as well as between grain boundaries and grain interiors. PEEM in UHV enabled us to identify a crucial intermediate step toward modifying grain boundary properties in CdCl$_2$-treated CdTe, namely, air exposure. Our results demonstrate that PEEM is a valuable tool to image secondary junctions in CdTe (e.g. from grain-to-grain or grain boundary-to-grain) that impact local junction formation and carrier separation and collection.

## Acknowledgement

This work was supported by a U.S. Department of Energy, Office of Energy Efficiency and Renewable Energy SunShot Initiative BRIDGE award (DE-FOA-0000654 CPS25859), the Center for Integrated Nanotechnologies, an Office of Science User Facility (DE-AC04-94AL85000), a National Science Foundation PFI:AIR-RA:Advanced Thin-Film Photovoltaics for Sustainable Energy award (1538733), and Sandia Laboratory Directed Research and Development (LDRD). Sandia National Laboratories is a multi-mission laboratory managed and operated by National Technology and Engineering Solutions of Sandia, LLC., a wholly owned subsidiary of Honeywell International, Inc., for the U.S. Department of Energy's National Nuclear Security Administration under contract DE-NA0003525.

## References

[1] H. R. Moutinho, M. M. Al-Jassim, D. H. Levi, P. C. Dippo, and L. L. Kazmerski, "Effects of CdCl$_2$ treatment on the recrystallization and electro-optical properties of CdTe thin films," *Journal of Vacuum Science & Technology A*, vol. 16, pp. 1251-1257, 1998.

[2] I. Visoly-Fisher, S. R. Cohen, K. Gartsman, A. Ruzin, and D. Cahen, "Understanding the Beneficial Role of Grain Boundaries in Polycrystalline Solar Cells from Single-Grain-Boundary Scanning Probe Microscopy," *Advanced Functional Materials*, vol. 16, pp. 649–660, 2006.

[3] C.-S. Jiang, H. R. Moutinho, R. G. Dhere, and M. M. Al-Jassim, "The Nanometer-Resolution Local Electrical Potential and Resistance Mapping of CdTe Thin Films," *IEEE Journal of Photovoltaics*, vol. 3, NO. 4, pp. 1383-1388, 2013.

[4] C. Li, Y. Wu, J. Poplawsky, T. J. Pennycook, N. Paudel, W. Yin, S. J. Haigh, M. P. Oxley, A. R. Lupini, M. Al-Jassim, S. J. Pennycook, and Y. Yan, "Grain-Boundary-Enhanced Carrier Collection in CdTe Solar Cells", *Physics Review Letters*, vol. 112, 156103, 2014.

[5] E. S. Barnard, B. Ursprung, E. Colegrove, H. R. Moutinho, N. J. Borys, B. E. Hardin, C. H. Peters, W. K. Metzger, and P. J. Schuck, "3D Lifetime Tomography Reveals How CdCl$_2$ Improves Recombination Throughout CdTe Solar Cells," *Advanced Materials*, vol. 29, 1603801, 2017.

[6] J. D. Major, M. Al Turkestani, L. Bowen, M. Brossard, C. Li, P. Lagoudakis, S. J. Pennycook, L. J. Phillips, R. E. Treharne, and K. Durose, "In-depth analysis of chloride treatments for thin-film CdTe solar cells," *Nature Communications*, vol. 7, 13231, 2016.

[7] D. Hanks, M. Weir, K. Horsley, T. Hofmann, L. Weinhardt, M. Bär, K. Barricklow, P. Kobyakov, W. Sampath, and C. Heske, "Photoemission Study of CdTe Surfaces After Low-Energy Ion Treatments," *38th IEEE Phot. Spec. Conf.*, pp. 396-399, 2011.

[8] J. M. Kephart, R. M. Geisthardt, and W. S. Sampath, "Optimization of CdTe thin-film solar cell efficiency using a sputtered, oxygenated CdS window layer," *Progress in Photovoltaics*, vol. 23, pp. 1484-1492, 2016.

[9] D. E. Swanson, R. M. Geisthardt, J. T. McGoffin, J. D. Williams and J. R. Sites, "Improved CdTe Solar-Cell Performance by Plasma Cleaning the TCO Layer," *IEEE Journal of Photovoltaics*, vol. 3, no. 2, pp. 838-842, 2013.

[10] J. D. Major, L. Bowen, R. E. Treharne, L. J. Phillips, and K. Durose, "NH$_4$Cl Alternative to the CdCl$_2$ Treatment Step for CdTe Thin-Film Solar Cells," *IEEE Journal of Photovoltaics*, vol. 5, no. 1, pp. 386-389, 2015.

[11] J. D. Major, "Grain boundaries in CdTe thin film solar cells: a review," *Semiconductor Science and Technology*, vol. 31, 093001, 2016.

[12] J. Luria, Y. Kutes, A. Moore, L. Zhang, E. A. Stach, and B. D. Huey, "Charge transport in CdTe solar cells revealed by conductive tomographic atomic force microscopy," *Nature Energy*, vol. 1, 16150, 2016.

[13] J. Quadros, A. L. Pinto, H. R. Moutinho, R. G. Dhere, and L. R. Cruz, "Microtexture of chloride treated CdTe thin films deposited by CSS technique," *Journal of Material Science*, vol. 43, pp. 573-579, 2008.

[14] C.S Ferekides, D Marinskiy, V Viswanathan, B Tetali, V Palekis, P Selvaraj, and D.L Morel, "High efficiency CSS CdTe solar cells," *Thin Solid Films*, vol. 361–362, pp. 520–526, 2000.

[15] S. A. Jensen, J. M. Burst, J. N. Duenow, H. L. Guthrey, J. Moseley, H. R. Moutinho, S. W. Johnston, A. Kanevce, M. M. Al-Jassim, and W. K. Metzger, "Long carrier lifetimes in large-grain polycrystalline CdTe without CdCl$_2$," *Applied Physics Letters*, vol. 108, 263903, 2016.